\documentclass[twocolumn,showpacs,prl,aps,superscriptaddress,floatfix]{revtex4-1}
\usepackage{amsfonts}
\usepackage{mathrsfs}
\usepackage{amssymb}
\usepackage{bm}
\usepackage{graphicx}
\usepackage[colorlinks=true,linkcolor=blue,citecolor=red,urlcolor=blue]{hyperref}

\begin{document}

\title{Analytical study of quadratic and non-quadratic \\short-time behavior of quantum decay}
\author{Sergio Cordero}%
\email{cordero@fis.unam.mx}
\affiliation{Instituto de Ciencias F\'isicas, Universidad Nacional Aut\'onoma de M\'exico,  Apartado postal 48--3,
Cuernavaca, Morelos 62251, Mexico}
\author{Gast\'on Garc\'{i}a--Calder\'{o}n}
\email{gaston@fisica.unam.mx}
\affiliation{Instituto de F\'{i}sica, Universidad Nacional Aut\'{o}noma de M\'{e}xico, Apartado postal 20--364,
M\'{e}xico 01000, Distrito Federal, Mexico}

\date{\today}

\begin{abstract}
The short-time behavior of quantum decay of an unstable state initially located within an interaction region of finite range is investigated using a resonant expansion of the survival amplitude. It is shown that in general the short-time behavior of the
survival probability $S(t)$ has a dependence on the initial state and may behave either as $S(t)=1-\mathcal{O}(t^{3/2})$ or as $S(t)=1-\mathcal{O}(t^{2})$. The above cases are illustrated by solvable models. The experiment reported in Ref. \cite{raizen97} does not distinguish between the above short-time behaviors.
\end{abstract}


\maketitle

%
%\section{Introduction.}
%

\textit{Introduction.} The decay of unstable systems, corresponding to particle emission by tunneling out of a potential, has been a subject of attention since the early days of quantum mechanics. In 1928 Gamow  derived an expression for the exponential decay law and introduced the notion of the lifetime \cite{gamow28}, which provides the time scale for exponential decay and sets the meaning for short or long times in decay. In the fifties of last century, Khalfin  demonstrated  that if the energy spectra of the system is bounded by below, the exponential decay law cannot hold at long times \cite{khalfin58}.  At short times there is also a departure from the exponential decay behavior which, however, it is related to the energy moments of the Hamiltonian to the system \cite{khalfin68,sudarshan77,gaemers88,muga96}. It is usually \textit{assumed} that decay at short-times exhibits a quadratic behavior with time \cite{khalfin68,sudarshan77,ghirardi79,peres80,muga96,pascazio08}. The relevant quantity is the survival probability $S(t)=|A(t)|^2$, with $A(t)=\langle0|\exp (-i\mathbf{H}t/\hbar)|0\rangle$, that yields the probability that at time $t$ the system remains in the normalized initial state $|0\rangle$. Notice that expanding $\exp (-i\mathbf{H}t/\hbar)$  yields
\begin{eqnarray}
A(t) &=& 1-i\langle 0|\mathbf{H}|0\rangle t/\hbar - 2\langle 0|\mathbf{H}^2|0\rangle t^2/\hbar^2+ \nonumber \\ [.3cm]
&&\left \langle 0\left |\sum_{j=3}^\infty \frac{(-i\mathbf{H}t/\hbar)^j}{j!}\right |0 \right \rangle,
\label{uexp}
\end{eqnarray}
which leads to an expansion of $S(t)$ that involves only even powers of $t$. A quadratic behavior requires that at least the first two moments of the Hamiltonian to the system are finite. The experimental verification of the short-time behavior of decay was provided some years ago and seems to be consistent with an initial quadratic behavior \cite{raizen97}. However, from a theoretical point of view, it is not obvious that the series expansion of $S(t)$ mentioned above converges or even that it is defined. In this context. there are problems that have been rarely explored as the conditions that may lead to a non-quadratic behavior at short times \cite{muga96,gcrr01}. Some recent work has discussed a $t^{3/2}$ short-time behavior of decay in the context of specific models \cite{muga96b,marchewka00,granot09,sokolovski12}. It is also worth mentioning that a $t^{3/2}$ short-time behavior has been found in studies involving transients in non-decay problems \cite{granot10}.

Here we consider an approach to the time evolution of decay based on a resonant expansion of the survival amplitude  that has been studied intensively for the exponential and long-time regimes \cite{gcmm95,gc10,gc11}. The occurrence, however, of a double sum in the expression for $S(t)$, which in general does not commute, prevented its application to the discussion of the short-time behavior of decay \cite{gc92}. Recently, however, motivated by the considerations mentioned in the previous paragraph, we believe that we have found a way to circumvent the above situation.

In this work we address a rigorous investigation of the short-time behavior of decay for unstable systems. We obtain general conditions on the initial states so that $S(t)$ may exhibit either a time dependence as $t^{3/2}$ or as $t^2$.  We  also indicate that the experiment on the short-time decay behavior reported in Ref. \cite{raizen97} does not distinguish between the above two cases.

%
%\section{Formalism}
%
\textit{Resonant expansion.} We consider a simple yet no trivial description of the decay process that involves real potentials of arbitrary shape that vanish beyond the interval $(0,L)$, which is well justified since most effective potentials in physics are of short-range, and initial states that are confined initially within the interaction region. The above conditions are commonly found in quantum systems designed artificially, as low temperature multibarrier resonant tunneling structures \cite{sakaki87} or ultracold atoms confined in optical traps \cite{jochim11}. A relevant feature of these systems is that the decay process is essentially coherent (elastic). One may then exploit the analytical properties of the outgoing Green's function to the problem on the whole complex wave number plane where it possesses an infinite number of poles. These poles are in general simple and are distributed  in a well known manner \cite{newtonchap12}. This has led to a formulation of the time evolution of decay in terms of a purely discrete expansion that involves the residues (resonant states) at the poles of the outgoing Green's function to the problem \cite{gcmm95,gc10,gc11}. The resonant states $u_n(x)$ satisfy the Schr\"odinger equation of the problem obeying purely outgoing boundary conditions and hence  they also include the bound and antibound  states of the problem.

It is worth stressing that the resonant state formulation yields exactly the same results as a calculation using continuum states \cite{gcmv12}.

One should mention that the above analytical properties of the  the outgoing Green's function remain valid for potentials having tails that go faster than an exponential at infinity, as for example having Gaussian tails \cite{newtonchap12}. The outgoing Green's function  for potentials having exponential tails may be extended analytically only through a finite region of the complex $k$ plane, and hence the corresponding expansion will consist in addition to a discrete pole expansion of an integral  contribution involving continuum of states. We believe, however, that this issue is mostly of mathematical interest since as pointed out above, most effective potentials in physics are negligible small after a distance and hence are beyond experimental scrutiny.

The survival amplitude may be expanded in terms of resonant states as \cite{gcmm95,gc10,gc11}
\begin{equation}
A(t)=\frac{1}{2}\sum_{n=-\infty}^{\infty} C_n\bar{C_n} \omega(iy_n),
\label{1e}
\end{equation}
where $\omega(iy_n)$ refers to the Faddeyeva function \cite{abramowitzchap7} with $y_n=-\exp(-i \pi /4)(\hbar/2m)^{1/2}\,\kappa_nt^{1/2}$,
and $\kappa_n=\alpha_n-i\beta_n$ relates to the complex energy eigenvalue $E_n=\hbar^2 \kappa_n^2/2m$. Notice that for bound states $\kappa_n=i\gamma_n$ with $\gamma_n >0$, and similarly for antibound states with $\gamma_n <0$.
The coefficients $C_n$ and $\bar{C_n}$ in (\ref{1e}) are
\begin{equation}
C_{n}=\int_{0}^{L}\psi \left( x,0\right) u_{n}(x)\,dx;\,
\bar{C}_{n}=\int_{0}^{L}\psi ^{\ast }\left( x,0\right)u_{n}(x)\,dx.
\label{4}
\end{equation}
The above coefficients fulfill the relationship \cite{gc10,gc11}
\begin{equation}
\frac{1}{2}\sum_{n=-\infty}^{\infty}\,C_n\bar{C}_n =1
\label{5}
\end{equation}
and the sum rules
\begin{equation}
\sum_{n=-\infty}^{\infty}\,{C_n\bar{C}_n \kappa_n}=0,
\label{6}
\end{equation}
and
\begin{equation}
\sum_{n=-\infty}^{\infty}\,\frac{C_n\bar{C}_n}{\kappa_n}=0.
\label{6a}
\end{equation}
Notice that Eq. (\ref{1e}) follows using
\begin{equation}
\psi(x,t)=(1/2)\sum_{n=-\infty}^{\infty}C_nu_n(x)\omega(iy_n),
\label{expsi}
\end{equation}
where $0 \leq x \leq L$ \cite{gcmm95,gc10,gc11}. Since $\omega(0)=1$ \cite{abramowitzchap7}, then $\psi(x,0)=(1/2)\sum_{n=-\infty}^{\infty}C_nu_n(x)$. Using that $\mathbf{H}u_n(x)=E_nu_n(x)$ and  the definition of $\bar{C}_n$  given in (\ref{4}), allows to express the first moment of $\mathbf{H}$ as
\begin{equation}
\langle 0|\mathbf{H}| 0 \rangle  \equiv \langle \mathbf{H} \rangle = \frac{1}{2}\left (\frac{\hbar^2}{2m}\right ) \sum_{n=-\infty}^{\infty}C_n\bar{C}_n\kappa^2_n,
\label{1stmoment}
\end{equation}
which is a finite quantity as follows by inspection of the conditions satisfied by the potential and the initial wave function.

The function $\omega(iy_n)$, which may be evaluated by well developed numerical methods \cite{poppe90}, may be written as
the convergent expansion (for any value of $iy_n$) \cite{abramowitzchap7},
\begin{equation}
\omega(iy_n) = \sum_{s=0}^{\infty} \frac{(a\kappa_nt^{1/2})^s}{\Gamma(1+s/2)},\quad a=e^{-i\pi/4} \left ({\hbar/2m}\right )^{1/2}.
\label{7}
\end{equation}
One may write, therefore, $\omega(iy_n)$  at short times as $\omega(iy_n)= 1+a\kappa_nt^{1/2}+a^2\kappa_n^2t + \dots $. Substitution of this expression into (\ref{1e}) allows to write $A(t)$ at short times as
\begin{eqnarray}
&&A(t)= \frac{1}{2}\sum_{n=-\infty}^{\infty}C_n\bar{C}_n +\frac{a}{2}\sum_{n=-\infty}^{\infty}C_n\bar{C}_n\kappa_n t^{1/2} +\nonumber \\ [.3cm]
&& \frac{a^2}{2}\sum_{n=-\infty}^{\infty}C_n\bar{C}_n\kappa_n^2t + \frac{1}{2}\sum_{n=-\infty}^{\infty}C_n\bar{C}_n \sum_{s=3}^{\infty} \frac{(a\kappa_nt^{1/2})^s}{\Gamma(1+s/2)}.
\label{1exp}
\end{eqnarray}

The term with $s=3$ in Eq. (\ref{1exp}) reads
\begin{equation}
\frac{a^3}{2\Gamma(5/2)} \left (\sum_{n=-\infty}^\infty {\cal C}_n\bar{\cal C}_n\kappa_n^3 \right) t^{3/2}.
\label{12}
\end{equation}
However, depending on the characteristics of the initial state $\psi(x,0)$, the sum in (\ref{12}) may vanish, be a constant or diverge. Let us first analyze the case where it vanishes. Then, we may write (\ref{1exp}) with the sum over $s$ starting from $s=4$ in the form
\begin{equation}
A(t)= 1-\frac{i}{\hbar}\langle \mathbf{H}\rangle t + \frac{1}{2}\sum_{n=-\infty}^{\infty}C_n\bar{C}_n\mathcal{O}_n(t^{2}),
\label{A.short}
\end{equation}
where we have used, respectively, Eqs. (\ref{5}), (\ref{6}), (\ref{1stmoment}) and
$\mathcal{O}_n(t^{2})$,  with $\mathcal{O}$ the $O$-symbol \cite{bruijn}, expresses the fact that as $t \to 0$, the leading term in the remaining absolutely convergent sum over $s$  is $t^{2}$. Hence the survival probability may be written as
\begin{equation}
S(t) = 1 +\frac{1}{\hbar^2} \langle \mathbf{H}\rangle^2 t^2 - \mathcal{O}(t^r),
\label{Str}
\end{equation}
Since as $t \to 0$, $\mathcal{O}_n(t^{2})/t^{\nu} \to 0$, requires that  $\nu < 2$, it follows  that $\mathcal{O}(t^r)/t^{\nu} \to 0$ provided $r \geq 2$. Notice however that since $S(0)=1$, the decay process implies  $S(t) < 1$ for $t >0$. Hence, in order to avoid that the term proportional to $t^2$ in (\ref{Str}) yields an unphysical  time interval where $S(t) > 1$, necessarily $r=2$. This guarantees that $S(t)$ diminishes with time and hence
\begin{equation}
S(t)= 1-\mathcal{O} (t^{2}).
\label{ssquare}
\end{equation}
The above result seems to hold independently of whether or not the second moment $\langle \mathbf{H}^2\rangle \propto \sum_{n=-\infty}^{\infty}C_n\bar{C}_n\kappa_n^4$ is finite;
the  second case, where the sum in (\ref{12}) is a constant, gives that the leading term of $S(t)$ at short times is $t^{3/2}$, and the last case, where the sum in (\ref{12}) diverges, implies that such a term cannot be extracted from the sum over $s$ in (\ref{1exp}),
\textit{i.e.}, for each value of $n$ one has to perform the convergent sum over $s$,  and hence
\begin{equation}
A(t)= 1-\frac{i}{\hbar}\langle \mathbf{H}\rangle t + \frac{1}{2}\sum_{n=-\infty}^{\infty}C_n\bar{C}_n\mathcal{O}_n(t^{3/2}).
\label{A.shortfrac}
\end{equation}
Notice that otherwise it would be a contradiction with the argument leading to Eq. (\ref{ssquare}), that rests on the assumption that Eq. (\ref{12}) vanishes exactly. Hence necessarily $r=3/2$, and therefore as $t \to 0$,
\begin{equation}
S(t)= 1- \mathcal{O}(t^{3/2}).
\label{sfractional}
\end{equation}

It is worth mentioning here that in Ref. \cite{muga96} reports the possibility of a $t^{3/2}$ short-time dependence of $S(t)$ provided the second energy moment diverges. Though  reference to a pole expansion of $A(t)$ involving $\omega$-functions is made to account for a possible fractional behavior, the analysis there is actually based on the finiteness or not of the expressions
\begin{equation}
[\dot{A}(t)]_{t=0} = -(i/\hbar) \langle \mathbf{H} \rangle
\label{adot}
\end{equation}
and
\begin{equation}
[\ddot{A}(t)]_{t=0} = -(1/\hbar^2) \langle \mathbf{H}^2 \rangle,
\label{addot}
\end{equation}
the dot indicating derivative with respect to time, which are obtained from the series expansion of $\exp(-i\mathbf{H}t/\hbar)$. Using the exact expansion (\ref{1e}) one sees that $[\dot{A}(t)]_{t=0}$ remains as given by (\ref{adot}) above, but
\begin{equation}
\ddot{A}(t) = a^4\sum_{n=-\infty}^{\infty} C_n{\bar C}_n \kappa_n^4 \omega(iy_n) + \frac{a^3}{\sqrt{\pi}}\sum_{n=-\infty}^{\infty}C_n{\bar C}_n \kappa_n^3 \frac{1}{t^{1/2}}
\label{Addot}
\end{equation}
is different. One sees therefore that  $[\ddot{A}(t)]_{t=0}$ diverges unless (\ref{12}) vanishes, independently of whether  $\langle \mathbf{H}^2\rangle$ is finite or not, contrary to the result given in Ref. \cite{muga96}. This indicates that the series expansion of $\exp(-i\mathbf{H}t/\hbar)$ may lead to misleading results.

The short-time expressions given by Eqs. (\ref{ssquare}) and (\ref{sfractional}) suggest to consider the expression for $S(t)$
\begin{equation}
S(t) \approx 1 - \left ( \frac{t}{\tau^*} \right )^{\vartheta},
\label{St.short}
\end{equation}
with parameters $\vartheta$ and $\tau^*$,  to \textbf{adjust}  the short-time behavior of exact calculations, using (\ref{1e}), or experiment.

%
%\section{Numerical Analysis}
%
\textit{Model.} As pointed out above, the distinction between the $t^{3/2}$ and $t^2$  short-time behavior of $S(t)$ depends on
the properties of the initial states. Theoretically, this necessarily leads to model calculations.
For simplicity, we consider as a model a double-barrier resonant tunneling nanostructure \cite{sakaki87,ferry} that extends from $0$ to $L=15\ nm$, with $b=5\ nm$ (barrier widths), $w = 5\ nm$ (well widths), $V = 0.23\ eV$ (barrier heights) and $m = 0.067m_e$ (effective electron mass). There exist  well developed procedures to obtain the set of resonant states $\{u_n(x)\}$ and complex poles $\{\kappa_n\}$ for a given problem \cite{cgc10a,gc11}. We choose two different types of initial states within the interaction potential region  $ 0\leq  x \leq L$: a cutoff Gaussian pulse
\begin{equation}
\psi(x)= (1/2\pi\sigma^2)^{1/4}\exp[-(x-x_0)^2/4\sigma^2]
\label{gaussian}
\end{equation}
centered at $x_0 = L/2$  with pulse width  $\sigma\ll w$, to guarantee that the effect of cutting-off the tails is negligible,  and a sinusoidal pulse:
\begin{equation}
\psi_j(x) = \sqrt{2/w}\sin\left[k_j\left(x-b\right)\right]
\label{sinusolidal}
\end{equation}
with $b\leq x \leq b+w$, and zero elsewhere, where $k_j = j\pi/w$ for a fixed integer value $j$.

The reason for the above choice of initial states is that, in addition to mathematical simplicity, one expects on physical grounds that the decaying particle is initially confined within the interaction region and hence that possible tails  beyond that region are negligible. Of course, one may envisage an initial state having large non-negligible external tails. In that case, as time evolves, part of the external portions of the initial state would head towards the internal region and would interfere with the  decaying part giving origin to a transient behavior.  We are not addressing such a possibility in this work though it might be of interest to investigate it.

In order to study numerically the behavior of the survival probability at short times, it is convenient to define the quantity
\begin{equation}
A_N(t) = \frac{1}{\mu_N(0)}\sum_{n=-N}^N {\cal C}_n\bar{\cal C}_n \omega(iy_n),
\label{suman}
\end{equation}
where
\begin{equation}
\mu_N(j) = \sum_{n=-N}^N {\cal C}_n\bar{\cal C}_n\kappa_n^j, \qquad j=0,1,2,\dots
\label{mu}
\end{equation}
Hence, the moments of the Hamiltonian may be written as $\langle \mathbf{H}^j \rangle = (1/2) (\hbar^2/2m)^j \lim_{N \to \infty}\,\mu_N(2j)$. Since $\mu_N(0)\to 2$ when $N\to\infty$, one finds that $S_N(t) = |A_N(t)|^2$ fulfills $\lim_{N\to\infty} S_N(t) \to S(t)$. Here, $S_N(0)=1$ for any value  $N>0$. Moreover, if for two values $N$ and $N'$ such that $N\gg N'$ the corresponding survival probability satisfies $S_N(t) = S_{N'}(t)$ in a time interval, that implies that both approximations yield the correct behavior of $S(t)$.
\begin{figure}
\begin{center}
\includegraphics[width=0.9\linewidth]{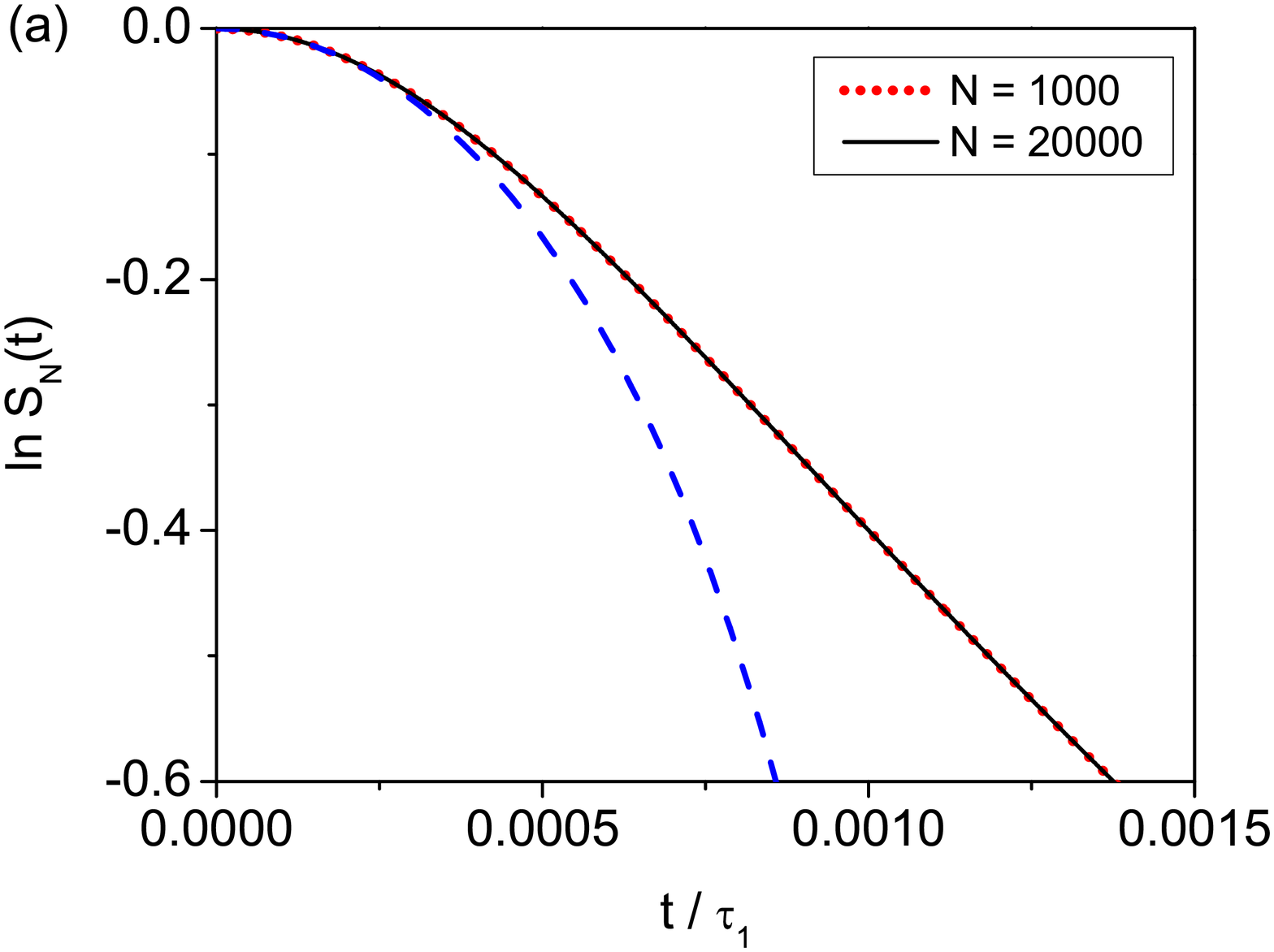}
\includegraphics[width=0.9\linewidth]{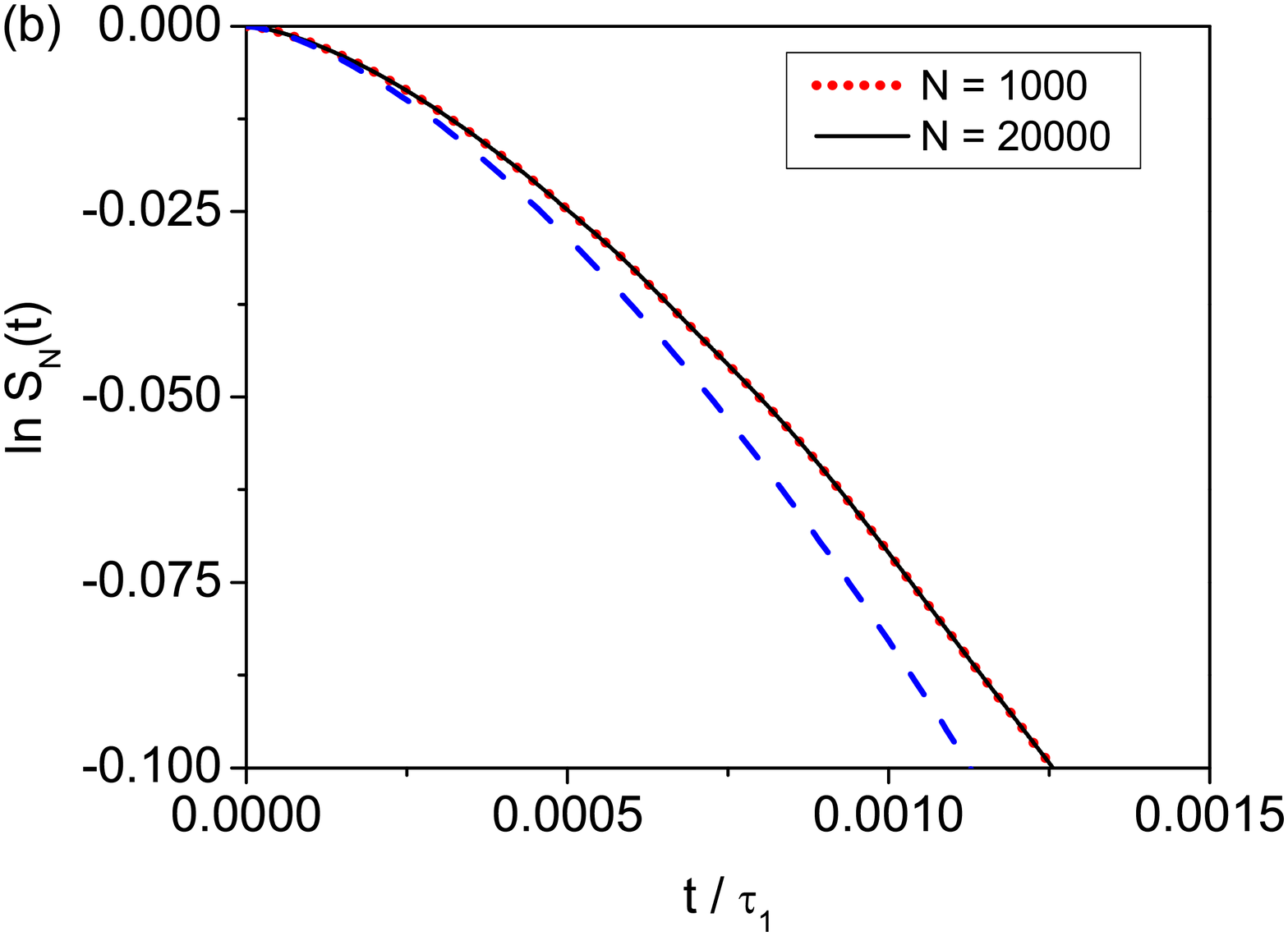}
\end{center}
\caption{(color online) Plot of $\ln S_N(t)$ as function time in lifetime units at short times for (a) An initial Gaussian pulse  and (b) An initial sinusoidal pulse. Both cases exhibit a comparison of the exact calculation (\ref{1e}), respectively, for $N=10^3$  (dotted line) and $N=2 \times 10^4$ poles (solid line). Each figure exhibits also the corresponding comparison with  (\ref{St.short}) (dashed line),  with adjustment parameters $\vartheta=2$ and $\tau^*=0.819$ fs in (a), and $\vartheta=3/2$ and $\tau^*=3.802$ fs in (b). See text.}
\label{f1}
\end{figure}

We have evaluated  Eq. (\ref{12}) for $N=10^3$ and $N=2 \times 10^4$ for both the Gaussian and the sinusoidal pulses and found that it vanishes for the Gaussian pulse and diverge for the sinusoidal pulse. This implies, according to our analysis above,  that the initial Gaussian state should produce a quadratic short-time behavior whereas the initial sinusoidal state a non-quadratic one. This needs to be confirmed by a comparison between an exact calculation, using (\ref{1e}), and the adjustment formula (\ref{St.short}).
We have also evaluated $\mu_N(j)$ for values $j=4,...,8$ for the above pulses and found that these quantities are finite for the Gaussian pulse and diverge for the sinusoidal one.

Figures \ref{f1}a  and  \ref{f1}b, exhibit, respectively, plots of  $\ln S_N(t)$ at short times in units of the lifetime $\tau_1$, for the initial Gaussian pulse with $\sigma =w/10$ and for the initial sinusoidal pulse with $j=1$, using the exact pole expansion (\ref{1e}) with $N=10^3$ poles (dotted line) and  $N=2\times10^4$ (solid line). One sees, in each figure, that these curves are indistinguishable from each other which indicates excellent convergence using $N=10^3$ poles. Each of the above figures also exhibits the results of the calculation  employing the \textbf{adjust} formula  (\ref{St.short}) employing, respectively,  the origin and two other points of the corresponding exact calculation. In Fig. \ref{f1}a the adjustment yields $\vartheta = 2$ and $\tau^*\approx 0.819$ fs (dashed line) which confirms the quadratic short-time behavior given by Eq. (\ref{ssquare}). In this case, we find that $\tau^* \approx \tau_Z$ where $\tau_Z$ is the Zeno time defined by $\tau_Z = \hbar/\Delta E$, with $\Delta E^2 = \langle \mathbf{H}^2\rangle - \langle \mathbf{H}\rangle^2$ \cite{pascazio08}. We have obtained similar results for initial Gaussian states having different values of the width $\sigma$ provided $\sigma \ll  w$. Similarly, in Fig. \ref{f1}b,  the adjustment (dashed line) yields $\vartheta = 3/2$ and $\tau^* =3.802$ fs which confirms the fractional $t^{3/2}$ short-time behavior given by Eq. (\ref{sfractional}). Similar results occur for other values of $j$.

It is worth emphasizing that if the first two moments of $\mathbf{H}$ exist in the expansion given by Eq. (\ref{uexp}), then consistency with the expansion given by Eq. (\ref{1e}), which in general involves both quadratic and non-quadratic powers of $t$, requires that the term proportional to $t^{3/2}$ should vanish, as indeed we have numerically corroborated.

\begin{figure}
\begin{center}
\includegraphics[width=0.9\linewidth]{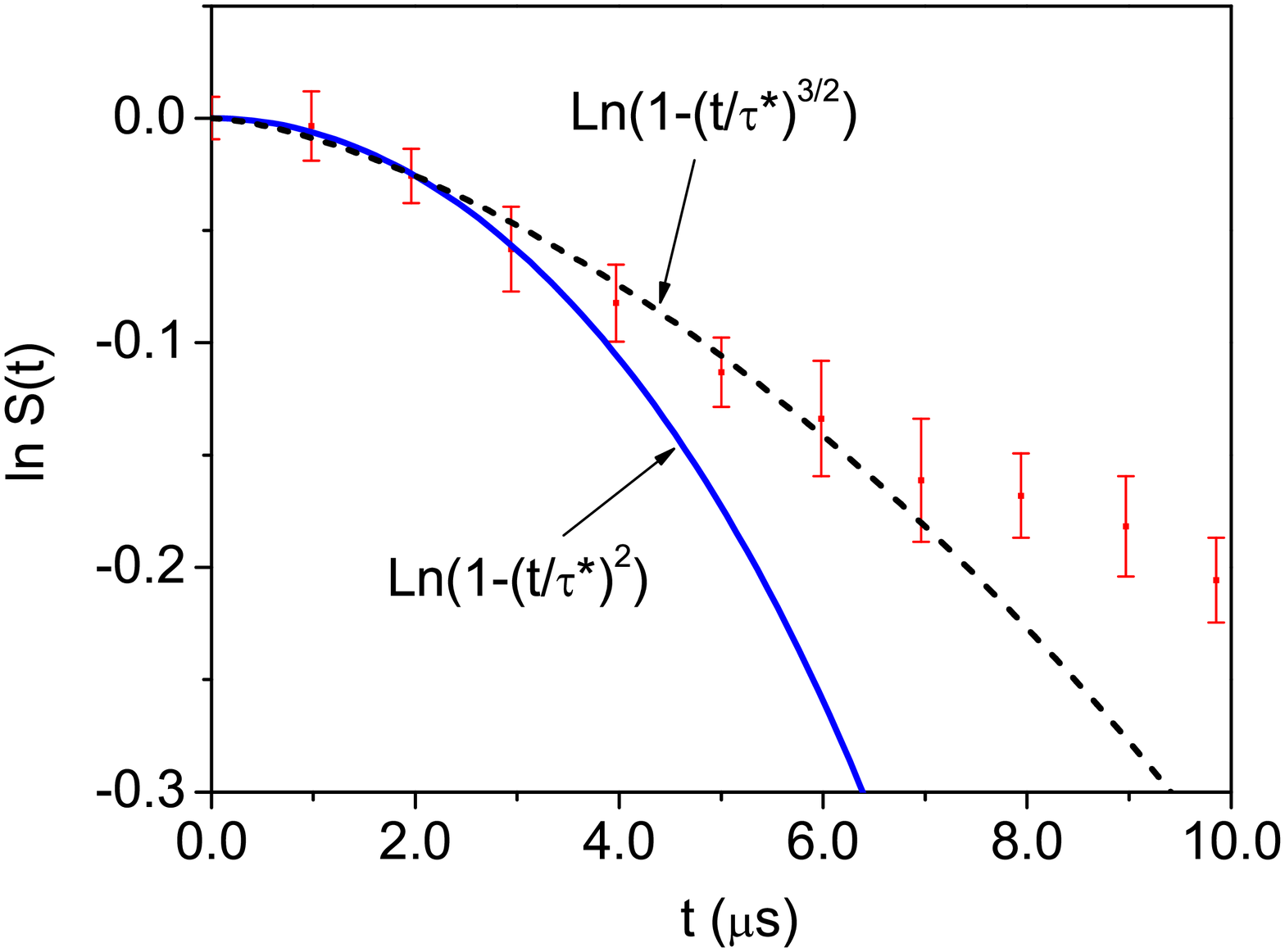}
\end{center}
\caption{(color online) Plot of the numerical adjustments at short times, using  Eq. (\ref{St.short}), to the corresponding experimental data taken from figure 3b of ref. \cite{raizen97}. See text.}
\label{f3}
\end{figure}

%
%Adjustment to experiment
%
\textit{Experiment.} The experiment of Ref. \cite{raizen97} involves an external potential that goes linearly with distance and hence our analysis is not strictly applicable. However, we find of interest to perform an elementary adjustment using (\ref{St.short}) to the  data given in Ref. \cite{raizen97}, which assumed a quadratic short-time behavior \cite{raizen97,raizen98}.  Since our analysis predicts that  the value of $\vartheta$ in Eq. (\ref{St.short}) is either $2$ or $3/2$, we need to consider only two experimental points to make the adjustment. We choose the points with a minimum error bar, in particular at $t=0$, and these correspond to Fig. 3b of Ref. \cite{raizen97}. In Fig. \ref{f3}, we plot the experimental data of Fig. 3b  at very short-times  and the corresponding adjustments using Eq. (\ref{St.short}) for $\vartheta=2$ and $\tau^*=12.55 \,\mu s$ (full line) and $\vartheta=1.5$ and $\tau^*=23.15\,\mu s$ (dotted line). We see that both short-time behaviors are consistent with experiment yet with a different value of the time scale $\tau^*$. May be a future experiment could discriminate between these two time scales.
%

%\section{Concluding Remarks}
%
\textit{Concluding remarks.} It is worth emphasizing that in general, the expansion of $A(t)$ in powers of $t$ is not defined. This means that the corresponding Taylor expansion around $t=0$ does not exist in general. The vanishing or not of Eq. (\ref{12}), which determines a quadratic or non-quadratic time evolution at short times, is very sensitive to the tails of the initial state, as exemplified by the Gaussian and sinusoidal initial states discussed here.  It is not clear, therefore, that initial physical states possess finite moments, a point that has been a subject of debate \cite{exner85sec}.  Further study on the characterization  of initial states is needed \cite{cgcrv11}. It is also worth mentioning that a non-quadratic  $t^{3/2}$ short-time behavior does not prevent the occurrence of the quantum Zeno effect \cite{sudarshan77,marchewka00,sokolovski12}. Finally, we believe that our results may be relevant for quests regarding the description of the short-time behavior of unstable systems in relativistic quantum field theory where it has been found that the second moment to the Hamiltonian diverges \cite{maiani98}.

\begin{acknowledgments}
S.C. acknowledges a post-doctoral fellowship from DGAPA-UNAM and  G.G-C. the partial financial support of DGAPA-UNAM under
grant IN103612.
\end{acknowledgments}

\begin{thebibliography}{34}%
\makeatletter
\providecommand \@ifxundefined [1]{%
 \@ifx{#1\undefined}
}%
\providecommand \@ifnum [1]{%
 \ifnum #1\expandafter \@firstoftwo
 \else \expandafter \@secondoftwo
 \fi
}%
\providecommand \@ifx [1]{%
 \ifx #1\expandafter \@firstoftwo
 \else \expandafter \@secondoftwo
 \fi
}%
\providecommand \natexlab [1]{#1}%
\providecommand \enquote  [1]{``#1''}%
\providecommand \bibnamefont  [1]{#1}%
\providecommand \bibfnamefont [1]{#1}%
\providecommand \citenamefont [1]{#1}%
\providecommand \href@noop [0]{\@secondoftwo}%
\providecommand \href [0]{\begingroup \@sanitize@url \@href}%
\providecommand \@href[1]{\@@startlink{#1}\@@href}%
\providecommand \@@href[1]{\endgroup#1\@@endlink}%
\providecommand \@sanitize@url [0]{\catcode `\\12\catcode `\$12\catcode
  `\&12\catcode `\#12\catcode `\^12\catcode `\_12\catcode `\%12\relax}%
\providecommand \@@startlink[1]{}%
\providecommand \@@endlink[0]{}%
\providecommand \url  [0]{\begingroup\@sanitize@url \@url }%
\providecommand \@url [1]{\endgroup\@href {#1}{\urlprefix }}%
\providecommand \urlprefix  [0]{URL }%
\providecommand \Eprint [0]{\href }%
\@ifxundefined \urlstyle {%
  \providecommand \doi  [0]{\begingroup \@sanitize@url \@doi}%
  \providecommand \@doi [1]{\endgroup \@@startlink {\doibase
  #1}doi:\discretionary {}{}{}#1\@@endlink }%
}{%
  \providecommand \doi  [0]{doi:\discretionary{}{}{}\begingroup
  \urlstyle{rm}\Url }%
}%
\providecommand \doibase [0]{http://dx.doi.org/}%
\providecommand \Doi [0]{\begingroup \@sanitize@url \@Doi }%
\providecommand \@Doi  [1]{\endgroup\@@startlink{\doibase#1}\@@Doi}%
\providecommand \@@Doi [1]{#1\@@endlink}%
\providecommand \selectlanguage [0]{\@gobble}%
\providecommand \bibinfo  [0]{\@secondoftwo}%
\providecommand \bibfield  [0]{\@secondoftwo}%
\providecommand \translation [1]{[#1]}%
\providecommand \BibitemOpen [0]{}%
\providecommand \bibitemStop [0]{}%
\providecommand \bibitemNoStop [0]{.\EOS\space}%
\providecommand \EOS [0]{\spacefactor3000\relax}%
\providecommand \BibitemShut  [1]{\csname bibitem#1\endcsname}%
%</preamble>
\bibitem [{\citenamefont {Wilkinson}\ \emph {et~al.}(1997)\citenamefont
  {Wilkinson}, \citenamefont {Bharucha}, \citenamefont {Fischer}, \citenamefont
  {Madison}, \citenamefont {Morrow}, \citenamefont {Niu}, \citenamefont
  {Sundaram},\ and\ \citenamefont {Raizen}}]{raizen97}%
  \BibitemOpen
  \bibfield  {author} {\bibinfo {author} {\bibfnamefont {S.~R.}\ \bibnamefont
  {Wilkinson}}, \bibinfo {author} {\bibfnamefont {C.~F.}\ \bibnamefont
  {Bharucha}}, \bibinfo {author} {\bibfnamefont {M.~C.}\ \bibnamefont
  {Fischer}}, \bibinfo {author} {\bibfnamefont {K.~W.}\ \bibnamefont
  {Madison}}, \bibinfo {author} {\bibfnamefont {P.~R.}\ \bibnamefont {Morrow}},
  \bibinfo {author} {\bibfnamefont {Q.}~\bibnamefont {Niu}}, \bibinfo {author}
  {\bibfnamefont {B.}~\bibnamefont {Sundaram}}, \ and\ \bibinfo {author}
  {\bibfnamefont {M.~G.}\ \bibnamefont {Raizen}},\ }
  {\bibfield  {journal} {\bibinfo  {journal} {Nature},\ }\textbf {\bibinfo
  {volume} {387}},\ \bibinfo {pages} {575} (\bibinfo {year}
  {1997})}\BibitemShut {NoStop}%
\bibitem [{\citenamefont {Gamow}(1928)}]{gamow28}%
  \BibitemOpen
  \bibfield  {author} {\bibinfo {author} {\bibfnamefont {G.}~\bibnamefont
  {Gamow}},\ }\href@noop {} {\bibfield  {journal} {\bibinfo  {journal} {Z.
  Phys.},\ }\textbf {\bibinfo {volume} {51}},\ \bibinfo {pages} {204} (\bibinfo
  {year} {1928})}\BibitemShut {NoStop}%
\bibitem [{\citenamefont {Khalfin}(1958)}]{khalfin58}%
  \BibitemOpen
  \bibfield  {author} {\bibinfo {author} {\bibfnamefont {L.~A.}\ \bibnamefont
  {Khalfin}},\ }\href@noop {} {\bibfield  {journal} {\bibinfo  {journal} {Sov.
  Phys.--JETP},\ }\textbf {\bibinfo {volume} {6}},\ \bibinfo {pages} {1053}
  (\bibinfo {year} {1958})}\BibitemShut {NoStop}%
\bibitem [{\citenamefont {Khalfin}(1968)}]{khalfin68}%
  \BibitemOpen
  \bibfield  {author} {\bibinfo {author} {\bibfnamefont {L.~A.}\ \bibnamefont
  {Khalfin}},\ }\href@noop {} {\bibfield  {journal} {\bibinfo  {journal} {JETP
  Lett.},\ }\textbf {\bibinfo {volume} {8}},\ \bibinfo {pages} {65} (\bibinfo
  {year} {1968})}\BibitemShut {NoStop}%
\bibitem [{\citenamefont {Chiu}\ and\ \citenamefont
  {Sudarshan}(1977)}]{sudarshan77}%
  \BibitemOpen
  \bibfield  {author} {\bibinfo {author} {\bibfnamefont {C.~B.}\ \bibnamefont
  {Chiu}}\ and\ \bibinfo {author} {\bibfnamefont {E.~C.~G.}\ \bibnamefont
  {Sudarshan}},\ }\href@noop {} {\bibfield  {journal} {\bibinfo  {journal}
  {Phys. Rev. D},\ }\textbf {\bibinfo {volume} {16}},\ \bibinfo {pages} {520}
  (\bibinfo {year} {1977})}\BibitemShut {NoStop}%
\bibitem [{\citenamefont {Gaemers}\ and\ \citenamefont
  {Visser}(1988)}]{gaemers88}%
  \BibitemOpen
  \bibfield  {author} {\bibinfo {author} {\bibfnamefont {K.}~\bibnamefont
  {Gaemers}}\ and\ \bibinfo {author} {\bibfnamefont {T.}~\bibnamefont
  {Visser}},\ } {\bibfield  {journal}
  {\bibinfo  {journal} {Physica A: Statistical Mechanics and its
  Applications},\ }\textbf {\bibinfo {volume} {153}},\ \bibinfo {pages} {234 }
  (\bibinfo {year} {1988})},\ ISSN \bibinfo {issn} {0378-4371}\BibitemShut
  {NoStop}%
\bibitem [{\citenamefont {Muga}\ \emph
  {et~al.}(1996){\natexlab{a}}\citenamefont {Muga}, \citenamefont {Wei},\ and\
  \citenamefont {Snider}}]{muga96}%
  \BibitemOpen
  \bibfield  {author} {\bibinfo {author} {\bibfnamefont {J.~G.}\ \bibnamefont
  {Muga}}, \bibinfo {author} {\bibfnamefont {G.~W.}\ \bibnamefont {Wei}}, \
  and\ \bibinfo {author} {\bibfnamefont {R.~F.}\ \bibnamefont {Snider}},\
  }\href@noop {} {\bibfield  {journal} {\bibinfo  {journal} {Europhys. Lett.},\
  }\textbf {\bibinfo {volume} {35}},\ \bibinfo {pages} {247} (\bibinfo {year}
  {1996}{\natexlab{a}})}\BibitemShut {NoStop}%
\bibitem [{\citenamefont {Ghirardi}\ \emph {et~al.}(1979)\citenamefont
  {Ghirardi}, \citenamefont {Omero}, \citenamefont {Weber},\ and\ \citenamefont
  {Rimini}}]{ghirardi79}%
  \BibitemOpen
  \bibfield  {author} {\bibinfo {author} {\bibfnamefont {G.~C.}\ \bibnamefont
  {Ghirardi}}, \bibinfo {author} {\bibfnamefont {C.}~\bibnamefont {Omero}},
  \bibinfo {author} {\bibfnamefont {T.}~\bibnamefont {Weber}}, \ and\ \bibinfo
  {author} {\bibfnamefont {A.}~\bibnamefont {Rimini}},\ }\href@noop {}
  {\bibfield  {journal} {\bibinfo  {journal} {Nuovo Cimento},\ }\textbf
  {\bibinfo {volume} {52 A}},\ \bibinfo {pages} {421} (\bibinfo {year}
  {1979})}\BibitemShut {NoStop}%
\bibitem [{\citenamefont {Peres}(1980)}]{peres80}%
  \BibitemOpen
  \bibfield  {author} {\bibinfo {author} {\bibfnamefont {A.}~\bibnamefont
  {Peres}},\ }\href@noop {} {\bibfield  {journal} {\bibinfo  {journal} {Ann. of
  Phys.},\ }\textbf {\bibinfo {volume} {129}},\ \bibinfo {pages} {33} (\bibinfo
  {year} {1980})}\BibitemShut {NoStop}%
\bibitem [{\citenamefont {Facchi}\ and\ \citenamefont
  {Pascazio}(2008)}]{pascazio08}%
  \BibitemOpen
  \bibfield  {author} {\bibinfo {author} {\bibfnamefont {P.}~\bibnamefont
  {Facchi}}\ and\ \bibinfo {author} {\bibfnamefont {S.}~\bibnamefont
  {Pascazio}},\ }\href@noop {} {\bibfield  {journal} {\bibinfo  {journal} {J.
  Phys. A: Math. Theor.},\ }\textbf {\bibinfo {volume} {41}},\ \bibinfo {pages}
  {493001} (\bibinfo {year} {2008})}\BibitemShut {NoStop}%
\bibitem [{\citenamefont {Garc\'\i{}a-Calder\'on}\ \emph
  {et~al.}(2001)\citenamefont {Garc\'\i{}a-Calder\'on}, \citenamefont
  {Riquer},\ and\ \citenamefont {Romo}}]{gcrr01}%
  \BibitemOpen
  \bibfield  {author} {\bibinfo {author} {\bibfnamefont {G.}~\bibnamefont
  {Garc\'\i{}a-Calder\'on}}, \bibinfo {author} {\bibfnamefont {V.}~\bibnamefont
  {Riquer}}, \ and\ \bibinfo {author} {\bibfnamefont {R.}~\bibnamefont
  {Romo}},\ }\href@noop {} {\bibfield  {journal} {\bibinfo  {journal} {J. Phys
  A: Math. Gen.},\ }\textbf {\bibinfo {volume} {34}},\ \bibinfo {pages} {4155}
  (\bibinfo {year} {2001})}\BibitemShut {NoStop}%
\bibitem [{\citenamefont {Muga}\ \emph
  {et~al.}(1996){\natexlab{b}}\citenamefont {Muga}, \citenamefont {Wei},\ and\
  \citenamefont {Snider}}]{muga96b}%
  \BibitemOpen
  \bibfield  {author} {\bibinfo {author} {\bibfnamefont {J.~G.}\ \bibnamefont
  {Muga}}, \bibinfo {author} {\bibfnamefont {G.~W.}\ \bibnamefont {Wei}}, \
  and\ \bibinfo {author} {\bibfnamefont {R.~F.}\ \bibnamefont {Snider}},\
  }\href@noop {} {\bibfield  {journal} {\bibinfo  {journal} {Ann. Phys.},\
  }\textbf {\bibinfo {volume} {252}},\ \bibinfo {pages} {336} (\bibinfo {year}
  {1996}{\natexlab{b}})}\BibitemShut {NoStop}%
\bibitem [{\citenamefont {Marchewka}\ and\ \citenamefont
  {Schuss}(2000)}]{marchewka00}%
  \BibitemOpen
  \bibfield  {author} {\bibinfo {author} {\bibfnamefont {A.}~\bibnamefont
  {Marchewka}}\ and\ \bibinfo {author} {\bibfnamefont {Z.}~\bibnamefont
  {Schuss}},\ } {\bibfield  {journal}
  {\bibinfo  {journal} {Phys. Rev. A},\ }\textbf {\bibinfo {volume} {61}},\
  \bibinfo {pages} {052107} (\bibinfo {year} {2000})}\BibitemShut {NoStop}%
\bibitem [{\citenamefont {Marchewka}\ and\ \citenamefont
  {Granot}(2009)}]{granot09}%
  \BibitemOpen
  \bibfield  {author} {\bibinfo {author} {\bibfnamefont {A.}~\bibnamefont
  {Marchewka}}\ and\ \bibinfo {author} {\bibfnamefont {E.}~\bibnamefont
  {Granot}},\ } {\bibfield  {journal}
  {\bibinfo  {journal} {Phys. Rev. A},\ }\textbf {\bibinfo {volume} {79}},\
  \bibinfo {pages} {012106} (\bibinfo {year} {2009})}\BibitemShut {NoStop}%
\bibitem [{\citenamefont {Sokolovski}\ \emph {et~al.}(2012)\citenamefont
  {Sokolovski}, \citenamefont {Pons},\ and\ \citenamefont
  {Kamalov}}]{sokolovski12}%
  \BibitemOpen
  \bibfield  {author} {\bibinfo {author} {\bibfnamefont {D.}~\bibnamefont
  {Sokolovski}}, \bibinfo {author} {\bibfnamefont {M.}~\bibnamefont {Pons}}, \
  and\ \bibinfo {author} {\bibfnamefont {T.}~\bibnamefont {Kamalov}},\ } {\bibfield  {journal} {\bibinfo  {journal}
  {Phys. Rev. A},\ }\textbf {\bibinfo {volume} {86}},\ \bibinfo {pages}
  {022110} (\bibinfo {year} {2012})}\BibitemShut {NoStop}%
\bibitem [{\citenamefont {Granot}\ and\ \citenamefont
  {Marchewka}(2010)}]{granot10}%
  \BibitemOpen
  \bibfield  {author} {\bibinfo {author} {\bibfnamefont {E.}~\bibnamefont
  {Granot}}\ and\ \bibinfo {author} {\bibfnamefont {A.}~\bibnamefont
  {Marchewka}},\ }\href@noop {} {\bibfield  {journal} {\bibinfo  {journal}
  {Phys. Rev. A},\ }\textbf {\bibinfo {volume} {81}},\ \bibinfo {pages}
  {032125} (\bibinfo {year} {2010})},\ \bibinfo {note} {and references
  therein}\BibitemShut {NoStop}%
\bibitem [{\citenamefont {Garc\'{i}a-Calder\'{o}n}\ \emph
  {et~al.}(1995)\citenamefont {Garc\'{i}a-Calder\'{o}n}, \citenamefont
  {Mateos},\ and\ \citenamefont {Moshinsky}}]{gcmm95}%
  \BibitemOpen
  \bibfield  {author} {\bibinfo {author} {\bibfnamefont {G.}~\bibnamefont
  {Garc\'{i}a-Calder\'{o}n}}, \bibinfo {author} {\bibfnamefont {J.~L.}\
  \bibnamefont {Mateos}}, \ and\ \bibinfo {author} {\bibfnamefont
  {M.}~\bibnamefont {Moshinsky}},\ }\href@noop {} {\bibfield  {journal}
  {\bibinfo  {journal} {Phys. Rev. Lett.},\ }\textbf {\bibinfo {volume} {74}},\
  \bibinfo {pages} {337} (\bibinfo {year} {1995})}\BibitemShut {NoStop}%
\bibitem [{\citenamefont {Garc\'{\i}a-Calder\'on}(2010)}]{gc10}%
  \BibitemOpen
  \bibfield  {author} {\bibinfo {author} {\bibfnamefont {G.}~\bibnamefont
  {Garc\'{\i}a-Calder\'on}},\ }\href@noop {} {\bibfield  {journal} {\bibinfo
  {journal} {Adv. Quant. Chem.},\ }\textbf {\bibinfo {volume} {60}},\ \bibinfo
  {pages} {407 } (\bibinfo {year} {2010})}\BibitemShut {NoStop}%
\bibitem [{\citenamefont {Garc\'\i{}a-Calder\'on}(2011)}]{gc11}%
  \BibitemOpen
  \bibfield  {author} {\bibinfo {author} {\bibfnamefont {G.}~\bibnamefont
  {Garc\'\i{}a-Calder\'on}},\ }\href@noop {} {\bibfield  {journal} {\bibinfo
  {journal} {AIP Conference Proceedings},\ }\textbf {\bibinfo {volume}
  {1334}},\ \bibinfo {pages} {84} (\bibinfo {year} {2011})}\BibitemShut
  {NoStop}%
\bibitem [{\citenamefont {Garc\'{i}a-Calder\'{o}n}(1992)}]{gc92}%
  \BibitemOpen
  \bibfield  {author} {\bibinfo {author} {\bibfnamefont {G.}~\bibnamefont
  {Garc\'{i}a-Calder\'{o}n}},\ }\enquote {\bibinfo {title} {Resonant states and
  the decay process: Symmetries in physics},}\ \ (\bibinfo  {publisher}
  {Springer--Verlag, Berlin},\ \bibinfo {year} {1992})\ Chap.~\bibinfo
  {chapter} {17}, pp.\ \bibinfo {pages} {252--272}\BibitemShut {NoStop}%
\bibitem [{\citenamefont {Tsuchiya}\ \emph {et~al.}(1987)\citenamefont
  {Tsuchiya}, \citenamefont {Matsusue},\ and\ \citenamefont
  {Sakaki}}]{sakaki87}%
  \BibitemOpen
  \bibfield  {author} {\bibinfo {author} {\bibfnamefont {M.}~\bibnamefont
  {Tsuchiya}}, \bibinfo {author} {\bibfnamefont {T.}~\bibnamefont {Matsusue}},
  \ and\ \bibinfo {author} {\bibfnamefont {H.}~\bibnamefont {Sakaki}},\
  }\href@noop {} {\bibfield  {journal} {\bibinfo  {journal} {Phys. Rev.
  Lett.},\ }\textbf {\bibinfo {volume} {59}},\ \bibinfo {pages} {2356}
  (\bibinfo {year} {1987})}\BibitemShut {NoStop}%
\bibitem [{\citenamefont {Serwane}\ \emph {et~al.}(2011)\citenamefont
  {Serwane}, \citenamefont {Z\"urn}, \citenamefont {Lompe}, \citenamefont
  {Ottenstein}, \citenamefont {Wenz},\ and\ \citenamefont
  {S.Jochim}}]{jochim11}%
  \BibitemOpen
  \bibfield  {author} {\bibinfo {author} {\bibfnamefont {F.}~\bibnamefont
  {Serwane}}, \bibinfo {author} {\bibfnamefont {G.}~\bibnamefont {Z\"urn}},
  \bibinfo {author} {\bibfnamefont {T.}~\bibnamefont {Lompe}}, \bibinfo
  {author} {\bibfnamefont {T.}~\bibnamefont {Ottenstein}}, \bibinfo {author}
  {\bibfnamefont {A.~N.}\ \bibnamefont {Wenz}}, \ and\ \bibinfo {author}
  {\bibnamefont {S.Jochim}},\ }\href@noop {} {\bibfield  {journal} {\bibinfo
  {journal} {Science},\ }\textbf {\bibinfo {volume} {332}},\ \bibinfo {pages}
  {336} (\bibinfo {year} {2011})}\BibitemShut {NoStop}%
\bibitem [{\citenamefont {Newton}(2002)}]{newtonchap12}%
  \BibitemOpen
  \bibfield  {author} {\bibinfo {author} {\bibfnamefont {R.~G.}\ \bibnamefont
  {Newton}},\ }\href@noop {} {\emph {\bibinfo {title} {Scattering Theory of
  Waves and Particles}}},\ \bibinfo {edition} {2nd}\ ed.\ (\bibinfo
  {publisher} {Dover Publications INC.},\ \bibinfo {year} {2002})\ \bibinfo
  {note} {chap. 12}\BibitemShut {NoStop}%
\bibitem [{\citenamefont {Garc\'\i{}a-Calder\'on}\ \emph
  {et~al.}()\citenamefont {Garc\'\i{}a-Calder\'on}, \citenamefont {M\'attar},\
  and\ \citenamefont {Villavicencio}}]{gcmv12}%
  \BibitemOpen
  \bibfield  {author} {\bibinfo {author} {\bibfnamefont {G.}~\bibnamefont
  {Garc\'\i{}a-Calder\'on}}, \bibinfo {author} {\bibfnamefont {A.}~\bibnamefont
  {M\'attar}}, \ and\ \bibinfo {author} {\bibfnamefont {J.}~\bibnamefont
  {Villavicencio}},\ }\href@noop {} {}\bibinfo {note}
  {ArXiv:1205.0487}\BibitemShut {NoStop}%
\bibitem [{\citenamefont {Abramowitz}\ and\ \citenamefont
  {Stegun}(1968)}]{abramowitzchap7}%
  \BibitemOpen
  \bibfield  {author} {\bibinfo {author} {\bibfnamefont {M.}~\bibnamefont
  {Abramowitz}}\ and\ \bibinfo {author} {\bibfnamefont {I.}~\bibnamefont
  {Stegun}},\ }\href@noop {} {\emph {\bibinfo {title} {Handbook of Mathematical
  Functions}}}\ (\bibinfo  {publisher} {Dover, N. Y.},\ \bibinfo {year}
  {1968})\ \bibinfo {note} {chap. 7}\BibitemShut {NoStop}%
\bibitem [{\citenamefont {Poppe}\ and\ \citenamefont {Wijers}(1990)}]{poppe90}%
  \BibitemOpen
  \bibfield  {author} {\bibinfo {author} {\bibfnamefont {G.~P.~M.}\
  \bibnamefont {Poppe}}\ and\ \bibinfo {author} {\bibfnamefont {C.~M.~J.}\
  \bibnamefont {Wijers}},\ }\href@noop {} {\bibfield  {journal} {\bibinfo
  {journal} {ACM Transactions on Mathematical Software},\ }\textbf {\bibinfo
  {volume} {16}},\ \bibinfo {pages} {38} (\bibinfo {year} {1990})}\BibitemShut
  {NoStop}%
\bibitem [{\citenamefont {{De Bruijn}}(1981)}]{bruijn}%
  \BibitemOpen
  \bibfield  {author} {\bibinfo {author} {\bibfnamefont {N.~G.}\ \bibnamefont
  {{De Bruijn}}},\ }\href@noop {} {\emph {\bibinfo {title} {Asymptotic Methods
  in Analysis}}}\ (\bibinfo  {publisher} {Dover Publications INC.},\ \bibinfo
  {year} {1981})\BibitemShut {NoStop}%
\bibitem [{\citenamefont {Ferry}\ and\ \citenamefont {Goodnick}(1997)}]{ferry}%
  \BibitemOpen
  \bibfield  {author} {\bibinfo {author} {\bibfnamefont {D.~K.}\ \bibnamefont
  {Ferry}}\ and\ \bibinfo {author} {\bibfnamefont {S.~M.}\ \bibnamefont
  {Goodnick}},\ }\href@noop {} {\emph {\bibinfo {title} {Transport in
  Nanostructures}}}\ (\bibinfo  {publisher} {Cambridge University Press, United
  Kingdom},\ \bibinfo {year} {1997})\ \bibinfo {note} {chap. 3}\BibitemShut
  {NoStop}%
\bibitem [{\citenamefont {Cordero}\ and\ \citenamefont
  {Garc\'{i}a-Calder\'{o}n}(2010)}]{cgc10a}%
  \BibitemOpen
  \bibfield  {author} {\bibinfo {author} {\bibfnamefont {S.}~\bibnamefont
  {Cordero}}\ and\ \bibinfo {author} {\bibfnamefont {G.}~\bibnamefont
  {Garc\'{i}a-Calder\'{o}n}},\ }\href@noop {} {\bibfield  {journal} {\bibinfo
  {journal} {J. Phys. A: Math. Theor.},\ }\textbf {\bibinfo {volume} {43}},\
  \bibinfo {pages} {185301} (\bibinfo {year} {2010})}\BibitemShut {NoStop}%
\bibitem [{\citenamefont {Niu}\ and\ \citenamefont {Raizen}(1998)}]{raizen98}%
  \BibitemOpen
  \bibfield  {author} {\bibinfo {author} {\bibfnamefont {Q.}~\bibnamefont
  {Niu}}\ and\ \bibinfo {author} {\bibfnamefont {M.~G.}\ \bibnamefont
  {Raizen}},\ }\href@noop {} {\bibfield  {journal} {\bibinfo  {journal} {Phys.
  Rev. Lett.},\ }\textbf {\bibinfo {volume} {80}},\ \bibinfo {pages} {3491}
  (\bibinfo {year} {1998})}\BibitemShut {NoStop}%
\bibitem [{\citenamefont {Exner}(1985)}]{exner85sec}%
  \BibitemOpen
  \bibfield  {author} {\bibinfo {author} {\bibfnamefont {P.}~\bibnamefont
  {Exner}},\ }\href@noop {} {\emph {\bibinfo {title} {Open Quantum Systems and
  Feynman Integrals}}}\ (\bibinfo  {publisher} {D. Reidel Publishing Company},\
  \bibinfo {year} {1985})\ \bibinfo {note} {{sec. 1.3}}\BibitemShut {NoStop}%
\bibitem [{\citenamefont {Cordero}\ \emph {et~al.}(2011)\citenamefont
  {Cordero}, \citenamefont {Garc\'\i{}a-Calder\'on}, \citenamefont {Romo},\
  and\ \citenamefont {Villavicencio}}]{cgcrv11}%
  \BibitemOpen
  \bibfield  {author} {\bibinfo {author} {\bibfnamefont {S.}~\bibnamefont
  {Cordero}}, \bibinfo {author} {\bibfnamefont {G.}~\bibnamefont
  {Garc\'\i{}a-Calder\'on}}, \bibinfo {author} {\bibfnamefont {R.}~\bibnamefont
  {Romo}}, \ and\ \bibinfo {author} {\bibfnamefont {J.}~\bibnamefont
  {Villavicencio}},\ }\href@noop {} {\bibfield  {journal} {\bibinfo  {journal}
  {Phys. Rev. A},\ }\textbf {\bibinfo {volume} {84}},\ \bibinfo {pages}
  {042118} (\bibinfo {year} {2011})}\BibitemShut {NoStop}%
\bibitem [{\citenamefont {Marinov}\ and\ \citenamefont
  {Segev}(1996)}]{segev96}%
  \BibitemOpen
  \bibfield  {author} {\bibinfo {author} {\bibfnamefont {M.~S.}\ \bibnamefont
  {Marinov}}\ and\ \bibinfo {author} {\bibfnamefont {B.}~\bibnamefont
  {Segev}},\ }\href@noop {} {\bibfield  {journal} {\bibinfo  {journal} {J.
  Phys. A: Math. Gen.},\ }\textbf {\bibinfo {volume} {29}},\ \bibinfo {pages}
  {2839} (\bibinfo {year} {1996})}\BibitemShut {NoStop}%
\bibitem [{\citenamefont {Maiani}\ and\ \citenamefont
  {Testa}(1998)}]{maiani98}%
  \BibitemOpen
  \bibfield  {author} {\bibinfo {author} {\bibfnamefont {L.}~\bibnamefont
  {Maiani}}\ and\ \bibinfo {author} {\bibfnamefont {M.}~\bibnamefont {Testa}},\
  }\href@noop {} {\bibfield  {journal} {\bibinfo  {journal} {Ann. Phys.},\
  }\textbf {\bibinfo {volume} {2003}},\ \bibinfo {pages} {353} (\bibinfo {year}
  {1998})}\BibitemShut {NoStop}%
\end{thebibliography}
\end{document}